\documentstyle[twoside,psfig,epsf]{article}

\catcode`\@=11
\long\def\@makefntext#1{
\protect\noindent \hbox to 3.2pt {\hskip-.9pt  
$^{{\eightrm\@thefnmark}}$\hfil}#1\hfill}		

\def\@makefnmark{\hbox to 0pt{$^{\@thefnmark}$\hss}}	
	
\def\ps@myheadings{\let\@mkboth\@gobbletwo
\def\@oddhead{\hfill\hbox{}\rightmark}   
\def\@oddfoot{}\def\@evenhead{\leftmark\hbox{}\hfill}\def\@evenfoot{}
\def\sectionmark##1{}\def\subsectionmark##1{}}

\oddsidemargin=\evensidemargin
\newcounter{sectionc}\newcounter{subsectionc}\newcounter{subsubsectionc}
\renewcommand{\section}[1] {\vspace{12pt}\addtocounter{sectionc}{1} 
\setcounter{subsectionc}{0}\setcounter{subsubsectionc}{0}\noindent 
	{\tenbf\thesectionc. #1}\par\vspace{5pt}}
\renewcommand{\subsection}[1] {\vspace{12pt}\addtocounter{subsectionc}{1} 
	\setcounter{subsubsectionc}{0}\noindent 
  {\bf\thesectionc.\thesubsectionc. {\kern1pt \bfit #1}}\par\vspace{5pt}}
\renewcommand{\subsubsection}[1] {\vspace{12pt}\addtocounter{subsubsectionc}{1}
	\noindent{\tenrm\thesectionc.\thesubsectionc.\thesubsubsectionc.
	{\kern1pt \tenit #1}}\par\vspace{5pt}}
\newcommand{\nonumsection}[1] {\vspace{12pt}\noindent{\tenbf #1}
	\par\vspace{5pt}}

\newcounter{appendixc}
\newcounter{subappendixc}[appendixc]
\newcounter{subsubappendixc}[subappendixc]
\renewcommand{\thesubappendixc}{\Alph{appendixc}.\arabic{subappendixc}}
\renewcommand{\thesubsubappendixc}
	{\Alph{appendixc}.\arabic{subappendixc}.\arabic{subsubappendixc}}

\renewcommand{\appendix}[1] {\vspace{12pt}
        \refstepcounter{appendixc}
        \setcounter{figure}{0}
        \setcounter{table}{0}
        \setcounter{lemma}{0}
        \setcounter{theorem}{0}
        \setcounter{corollary}{0}
        \setcounter{definition}{0}
        \setcounter{equation}{0}
        \renewcommand{\thefigure}{\Alph{appendixc}.\arabic{figure}}
        \renewcommand{\thetable}{\Alph{appendixc}.\arabic{table}}
        \renewcommand{\theappendixc}{\Alph{appendixc}}
        \renewcommand{\thelemma}{\Alph{appendixc}.\arabic{lemma}}
        \renewcommand{\thetheorem}{\Alph{appendixc}.\arabic{theorem}}
        \renewcommand{\thedefinition}{\Alph{appendixc}.\arabic{definition}}
        \renewcommand{\thecorollary}{\Alph{appendixc}.\arabic{corollary}}
        \renewcommand{\theequation}{\Alph{appendixc}.\arabic{equation}}
        \noindent{\tenbf Appendix#1}\par\vspace{5pt}}
\newcommand{\subappendix}[1] {\vspace{12pt}
        \refstepcounter{subappendixc}
        \noindent{\bf Appendix \thesubappendixc. {\kern1pt \bfit #1}}
	\par\vspace{5pt}}
\newcommand{\subsubappendix}[1] {\vspace{12pt}
        \refstepcounter{subsubappendixc}
        \noindent{\rm Appendix \thesubsubappendixc. {\kern1pt \tenit #1}}
	\par\vspace{5pt}}

\newcommand{\textlineskip}{\baselineskip=13pt}
\newcommand{\smalllineskip}{\baselineskip=10pt}

\newcommand{\copyrightheading}[1]
	{\vspace*{-2.5cm}\smalllineskip{\flushleft
	{\footnotesize International Journal of Theoretical and 
	 Applied Finance #1}\\
	{\footnotesize \copyright\kern2pt World Scientific Publishing
	 Company}\\
	 }}

\newcommand{\pub}[1]{{\begin{center}\footnotesize\smalllineskip 
	#1\\		
	\end{center}
	}}

\def\abstracts#1#2#3{{
	\centering{\begin{minipage}{4.5in}\footnotesize\baselineskip=10pt
	\parindent=0pt #1\par 
	\parindent=15pt #2\par
	\parindent=15pt #3
	\end{minipage}}\par}} 

\newcommand{\bibit}{\nineit}
\newcommand{\bibbf}{\ninebf}
\renewenvironment{thebibliography}[1]
	{\frenchspacing
	 \ninerm\baselineskip=11pt
	 \begin{list}{\arabic{enumi}.}
        {\usecounter{enumi}\setlength{\parsep}{0pt}     
	 \setlength{\leftmargin 12.7pt}{\rightmargin 0pt} 
         \setlength{\itemsep}{0pt} \settowidth
	{\labelwidth}{#1.}\sloppy}}{\end{list}}

\newcounter{itemlistc}
\newcounter{romanlistc}
\newcounter{alphlistc}
\newcounter{arabiclistc}

\newcommand{\fcaption}[1]{
        \refstepcounter{figure}
        \setbox\@tempboxa = \hbox{\footnotesize Fig.~\thefigure. #1}
        \ifdim \wd\@tempboxa > 5in
           {\begin{center}
        \parbox{5in}{\footnotesize\smalllineskip Fig.~\thefigure. #1}
            \end{center}}
        \else
             {\begin{center}
             {\footnotesize Fig.~\thefigure. #1}
              \end{center}}
        \fi}

\newcommand{\tcaption}[1]{
        \refstepcounter{table}
        \setbox\@tempboxa = \hbox{\footnotesize Table~\thetable. #1}
        \ifdim \wd\@tempboxa > 5in
           {\begin{center}
        \parbox{5in}{\footnotesize\smalllineskip Table~\thetable. #1}
            \end{center}}
        \else
             {\begin{center}
             {\footnotesize Table~\thetable. #1}
              \end{center}}
        \fi}

\def\@citex[#1]#2{\if@filesw\immediate\write\@auxout
	{\string\citation{#2}}\fi
\def\@citea{}\@cite{\@for\@citeb:=#2\do
	{\@citea\def\@citea{,}\@ifundefined
	{b@\@citeb}{{\bf ?}\@warning
	{Citation `\@citeb' on page \thepage \space undefined}}
	{\csname b@\@citeb\endcsname}}}{#1}}

\newif\if@cghi
\def\cite{\@cghitrue\@ifnextchar [{\@tempswatrue
	\@citex}{\@tempswafalse\@citex[]}}
\def\citelow{\@cghifalse\@ifnextchar [{\@tempswatrue
	\@citex}{\@tempswafalse\@citex[]}}
\def\@cite#1#2{{$\null^{#1}$\if@tempswa\typeout
	{IJCGA warning: optional citation argument 
	ignored: `#2'} \fi}}

\def\pmb#1{\setbox0=\hbox{#1}
	\kern-.025em\copy0\kern-\wd0
	\kern.05em\copy0\kern-\wd0
	\kern-.025em\raise.0433em\box0}

\def\fnt#1#2{\footnotetext{\kern-.3em
	{$^{\mbox{\scriptsize #1}}$}{#2}}}

\font\tenrm=cmr10
\font\tenit=cmti10 
\font\tenbf=cmbx10
\font\bfit=cmbxti10 at 10pt
\font\ninerm=cmr9
\font\nineit=cmti9
\font\ninebf=cmbx9
\font\eightrm=cmr8

\def\qed{\hbox{${\vcenter{\vbox{		 
   \hrule height 0.4pt\hbox{\vrule width 0.4pt height 6pt
   \kern5pt\vrule width 0.4pt}\hrule height 0.4pt}}}$}}

\def\theequation{\thesectionc.\arabic{equation}}  

    	{\setcounter{itemlistc}{0}		
	 \begin{list}{$\bullet$}		
	{\usecounter{itemlistc}			
	 \leftmargin10pt	       
	 \setlength{\parsep}{0pt}
	 \setlength{\itemsep}{0pt}     
	}}{\end{list}}

	{\setcounter{romanlistc}{0}		
	 \begin{list}{$($\roman{romanlistc}$)$}	
	{\usecounter{romanlistc}		
	 \leftmargin18pt 
	 \setlength{\parsep}{0pt}
	 \setlength{\itemsep}{0pt}	
	 \settowidth{\labelwidth}{#1}                          
	}}{\end{list}}

	{\setcounter{enumii}{0}			
	 \begin{list}{$($\alph{enumii}$)$}	
	{\usecounter{enumii}			
	 \leftmargin18pt		
	 \setlength{\parsep}{0pt}
	 \setlength{\itemsep}{0pt}	
	 \settowidth{\labelwidth}{#1}                          
	}}{\end{list}}

\newcommand{\cf}{{\it cf. }}
\newcommand{\var}{\mbox{Var}}
\newcommand{\average}[1]{E\left[#1\right]}

\begin{document}
\setlength{\textheight}{7.7truein}    

\normalsize\textlineskip
\thispagestyle{plain}
\setcounter{page}{1}



\centerline{\bf A CORRELATED STOCHASTIC VOLATILITY MODEL}
\baselineskip=13pt
\centerline{\bf MEASURING LEVERAGE AND OTHER STYLIZED FACTS}
\vspace*{0.37truein}

\centerline{\footnotesize JAUME MASOLIVER and JOSEP PERELL\'{O}}
\baselineskip=12pt
\centerline{\footnotesize\it Departament de F\'{\i}sica Fonamental. Universitat de Barcelona}
\baselineskip=10pt
\centerline{\footnotesize\it Diagonal, 647. E-08028 Barcelona, Spain}
\vspace*{0.225truein}
\pub{(\today.)}

\vspace*{0.21truein}
\abstracts{We present a stochastic volatility market model where volatility is correlated with return and  is represented by an Ornstein-Uhlenbeck process. With this model we exactly measure the leverage effect and other stylized facts, such as mean reversion, leptokurtosis and negative skewness. We also obtain a close analytical expression for the characteristic function and study the heavy tails of the probability distribution.}{}{}


\vspace*{4pt}
\normalsize\baselineskip=13pt   
\section{Introduction}	
\noindent
\looseness1 During decades the diffusion process known as the geometric Brownian motion has been widely accepted as one of the most universal models for speculative markets. It was proposed by Osborne in 1959~\cite{osborne} when he observed that empirical distributions of prices were biased and in disagreement with the theoretical distribution of Bachelier's arithmetic Brownian dynamics.

However, specially after the 1987 crash, the geometric Brownian motion and its subsequent Black-Scholes (B-S) formula were unable to reproduce the option price data of real markets. Several studies have collected empirical option prices in order to derive their implied volatility. These tests conclude that the implied volatility is not constant and it is well-fitted to a U-shaped function of  moneyness whose minimum is at moneyness near to 1, {\it i.e.}, when current stock price is equal to the striking price. This effect is known as the smile effect and shows the inadequacy of the Black-Scholes model since this assumes a constant volatility~\cite{jack}.

A possible way out to this inconsistency is assuming that volatility is not a constant but an unknown deterministic function of the underlying price. The deterministic volatility still allows to manage the option pricing within the B-S theory although in most of the cases it is not possible to derive an analytic option price. Within this approach, there exists the ARCH-GARCH models and their subsequent extensions~\cite{engle82,bolle}. These models do well in describing the implied volatility but their disadvantage is that some of their parameters substantially change with time frequency~\cite{engle-pat}.

The stochastic volatility (SV) models are another possible choice. These models assume the original log-Brownian motion model but, as their name indicates, with random volatility. At late eighties several different SV models were presented~\cite{hull87,scott87,wig87}. All of them propose a two-dimensional process involving two independent variables: the stock and the volatility. These works are basically interested on option pricing theory and ignore the statistical properties of the market model, although they are indeed able to reproduce the smile effect. 

After then there have appeared several papers extending and refining the original SV models but again many of them are solely interested on adequately describe empirical option prices. Stein and Stein~\cite{stein91} is an exception to this tendency because they study the most important statistical properties of a volatility model following an Ornstein-Uhlenbeck (O-U) process. 

We believe that the relative small number of works dealing with market dynamics based on SV models is essentially due to two reasons: (i) Their statistical properties are difficult to be analytically derived, and the analysis is even much more involved when there are correlations between volatility and stock. (ii) It is commonly asserted that empirical data available are not enough for obtaining a reliable estimation of all parameters involved in an SV model~\cite{fouquellibre}.

Our present work wants to modify these statements for the correlated Ornstein-Uhlenbeck stochastic volatility (O-U SV) model because we are able to analytically derive the main statistical properties of it. On the other hand, the leverage correlation recently observed by Bouchaud {\it etal.}~\cite{bouchPRL} allows us to estimate all parameters involved not only in our O-U SV process but eventually in any SV process. 

Recent research on empirical markets data has provided a set of requirements that a good market model must obey. In this regard Engle and Patton~\cite{engle-pat} have listed a number of stylized facts about the volatility. The results we will herein derive conclude that the SV models are good candidates fairly accomplishing these stylized facts. We will prove this and confront the statistical properties of the  correlated O-U SV model with the statistical properties of the daily return changes for the Dow-Jones stock index.

The paper is divided in 7 sections. After this introduction, we present our stochastic volatility market model in Section 2 and study the statistical properties of the volatility in Section 3. Section 4 is specifically focussed on the leverage effect. Section 5 is devoted to show how to estimate from the Dow-Jones index (1900-2000) the parameters of the model. Finally, Section 6 concentrates on the derivation of the probability distribution through obtaining the characteristic function. Conclusions are drawn in Section 7 and technical details are left to Appendices.

\section{The stochastic volatility market model\label{sec:ousv}}
\noindent
The starting point of any stochastic volatility model is the log-Brownian stochastic differential equation:
\begin{equation}
\frac{dS(t)}{S(t)}=\mu dt + \sigma dW_1(t),
\label{ds}
\end{equation}
where $\mu$ is the drift and $\sigma$ is the volatility. The SV models refine this dynamics taking $\sigma=\sigma(t)$ to be stochastic. There exist a large class of such models but, to our knowledge, the dynamics of $\sigma$ is not definitively associated to any specific process (see the monograph of Fouque {\it etal.}~\cite{fouquellibre} or the review by Ghysels {\it etal.}~\cite{ghys96}). We choose the Ornstein-Uhlenbeck (O-U) stochastic volatility model because, as we will see shortly, it is one of the simplest approaches still reproducing the main observed features of markets. We thus assume that the random dynamics of $\sigma(t)$ is given by~\cite{stein91}
\begin{equation}
d\sigma(t)=-\alpha(\sigma-\theta)dt+k dW_2(t).
\label{dsigma}
\end{equation}
Equations~(\ref{ds}) and~(\ref{dsigma}) contain a two-dimensional Wiener process $(W_1(t), W_2(t))$, where $dW_i(t)=\xi_i(t) dt$ $(i=1,2)$, and $\xi_{i}(t)$ is Gaussian white noise processes with zero mean, {\it i.e.},
\begin{equation}
\average{ \xi_i(t) }=0 \quad \mbox{and} \quad \average{ \xi_i(t) \xi_j(t')} = \rho_{ij} \delta(t-t').
\label{svxi}
\end{equation}
We note that the cross-correlation is given in terms of the Dirac delta function, {\it i.e.}, $\delta(x)=0$ for all $x\neq 0$ and 
\begin{equation}
\int_{-\infty}^{a} \phi(z) \delta(x-z) dz= 
\left\{
\begin{array}{cc}
\phi(x) & -\infty < x < a,
\\
0 & \mbox{otherwise;}
\end{array}
\right.
\label{diracG}
\end{equation}
where $\phi(x)$ is an arbitrary integrable function. Note that $\rho_{ij}=\rho_{ji}$ and $\rho_{ii}=1$, hence the components of $\rho_{ij}$ are reduced to
\begin{equation}
\rho_{ij}=
\left\{
\begin{array}{cc}
1 & \mbox{if } i=j \\
\rho & \mbox{if } i \neq j,
\end{array}
\right. 
\label{svrho}
\end{equation}
where the parameter $\rho$ is given by Eq.~(\ref{svxi}), {\it i.e.}, $\average{\xi_1(t)\xi_2(t')}=\rho\delta(t-t')$ which, in terms of the Wiener process, is equivalent to say that $\average{dW_1(t)dW_2(t')}=0$ when $t\neq t'$, and
$$
\average{dW_1(t)dW_2(t)}=\rho dt.
$$
Thus $\rho$ is the correlation coefficient and it has no definite sign ($-1 \leq \rho \leq 1$). However, it is known that a negative $\rho$ is able to provide the skewness observed in financial markets~\cite{fouquellibre}. One of our objectives here is not only to show that $\rho$ is negative but also to estimate its value from empirical data.

In what follows it turns to be more convenient to work with the zero-mean return defined as
\begin{equation}
dX\equiv\frac{dS}{S}-\mu dt,
\label{dXdef}
\end{equation}
and whose SDE reads
\begin{equation}
dX(t)=\sigma(t) dW_1(t).
\label{dx}
\end{equation}
The zero-mean return $X(t)$ has a simpler dynamics than the stock price $S(t)$ because it only contains the random fluctuation $\sigma dW_1$. Nevertheless, this process still retains the most interesting features of the whole dynamics. In the Appendix A, we give en explicit expression for $X(t)$ and derive some key features. 

\section{The Ornstein-Uhlenbeck volatility process\label{sec:ouvol}}
\noindent
We will now present the main properties of the O-U volatility. The starting point is the solution of Eq.~(\ref{dsigma}):
\begin{equation}
\sigma(t)=\sigma_0 e^{-\alpha (t-t_0)}+\theta (1-e^{-\alpha (t-t_0)})+k \int_{t_0}^{t} e^{-\alpha(t-t')} dW_2(t'),
\label{sigma}
\end{equation}
where we have assumed that the process started at time $t=t_0$, when volatility was $\sigma_0$. From now on we will assume that the volatility is in the stationary regime. This means that the market started long time ago thus $t_0\rightarrow -\infty$ and the stationary volatility reads
\begin{equation}
\sigma(t)=\theta+k \int_{-\infty}^{t} e^{-\alpha(t-t')} dW_2(t'),
\label{sigmastat}
\end{equation}
whose average value, variance and correlation are
\begin{equation}
\average{\sigma}= \theta,\qquad\var[\sigma]\equiv \average{\sigma^2}-\average{\sigma}^2= k^2/2\alpha,
\label{statave1}
\end{equation}
and
\begin{equation}
\average{\sigma(t+\tau)\sigma(t)}= \theta^2+(k^2/2\alpha) e^{-\alpha \tau}.
\label{corrsig}
\end{equation}
These expressions provide a physical interpretation of the parameters of the model, especially $\theta$, the expected volatility, and $\alpha$, the inverse of the volatility ``correlation time'' (see below). 

Note that the stationary volatility~(\ref{sigmastat}) is a linear functional of the Wiener process. Therefore, it is a Gaussian process (the same applies to ``transient" volatility~(\ref{sigma})). In consequence the stationary distribution is uniquely determined by the mean and variance given in Eq.~(\ref{statave1}). That is,
\begin{equation}
p_{\rm st}(\sigma)=\frac{1}{2\sqrt{\pi k^2/\alpha}}e^{-\alpha(\sigma-\theta)/k^2}.
\label{sigmapdfstat}
\end{equation}

Before proceeding further we want to address the question of the sign of $\sigma(t)$. If one adopts the O-U process~(\ref{dsigma}) as a model for stochastic volatility one may argue that $\sigma(t)$ has no definite sign which can be seen as an inconvenience for a ``good'' SV model. Let us see that this is not  really the case. First of all, the actual evaluation of volatility is very difficult, not to say impossible, since volatility itself is not observed. In practice, the so-called instantaneous volatility is derived from
\begin{equation}
\lim_{\Delta t \rightarrow 0} \sqrt{\left[X(t+\Delta t)-X(t)\right]^2/\Delta t},
\label{instvoldef}
\end{equation}
where we have used the zero-mean return defined above. Due to the limit $\Delta t\rightarrow 0$, this equation has to be taken as an infinitesimal difference. We thus define
\begin{equation}
\mbox{instantaneous volatility} \equiv \sqrt{dX(t)^2/dt}.
\label{instvoldef1}
\end{equation}
From Eq.~(\ref{dx}) and the fact $dW^2=dt$, we get
\begin{equation}
\mbox{instantaneous volatility}=\sqrt{\sigma(t)^2}=\left|\sigma(t)\right|.
\label{instvol}
\end{equation}
Observe that in this definition no sign is attached to the random variable $\sigma(t)$.

\subsection{The correlated process}
\noindent
As we have mentioned in Section 2, the model permits correlations between the stock and the volatility. We will now examine the important effects of these correlations. From Eq.~(\ref{sigmastat}), we see that the correlation between the stationary volatility and the random component of return variations, $dW_1(t)$, is
$$
\average{\sigma(t+\tau) dW_1(t)}=k\int_{-\infty}^{t+\tau} e^{-\alpha (t+\tau-t')} \average{dW_2(t')dW_1(t)},
$$
which, taking into account Eqs.~(\ref{svxi}) and (\ref{svrho}), can be written as
\begin{equation}
\average{\sigma(t+\tau)\frac{dW_1(t)}{dt}}=
\rho k\int_{-\infty}^{t+\tau}e^{-\alpha(t+\tau-t')}\delta(t-t')dt'.
\label{uncorr}
\end{equation}
Finally (\cf Eq.~(\ref{diracG}))
\begin{equation}
\average{\sigma(t+\tau) dW_1(t)}=
\left\{
\begin{array}{cc}
\rho k e^{-\alpha\tau} dt & \mbox{if } \tau>0, \\
0 & \mbox{if } \tau<0.
\end{array}
\right.
\label{sigmadW1}
\end{equation}
Therefore, for correlated SV processes {\it future volatility is correlated with past return variations, although past volatility and future return variations are completely uncorrelated}. Moreover, $\rho$ determines the sign of the correlation~(\ref{sigmadW1}) because $k$ is positive. In the next section we  will show how this property is able to reproduce the leverage effect observed in real markets. 

We finally note that, despite the existence of correlations, $\sigma(t)$ and $dW_1(t)$ are independent random quantities. This is a direct consequence of It\^o convention for stochastic integrals, because process $\sigma(t)$ is independent of its driving noise $dW_1(t)=\xi_1(t)dt$. Hence, as $\tau\rightarrow 0^-$ we have
\begin{equation}
\average{\sigma(t)dW_1(t)}=\average{\sigma(t)}\average{dW_1(t)}=0.
\label{mdx0}
\end{equation}
Since $dX(t)=\sigma(t)dW_1(t)$. Eq.~(\ref{mdx0}) implies that 
\begin{equation}
\average{dX(t)}=0,
\label{mdx}
\end{equation}
in accordance with the name ``zero-mean return'' given to $X(t)$. 

\subsection{Mean reversion\label{sec:mrev}}
\noindent
The effect of mean-reversion refers to the existence of a normal level of volatility to which volatility will eventually return. This effect can be observed in financial markets~\cite{ghys96}. Practitioners believe that the current volatility is high or low compared to a normal level of volatility and they assume that in the long run, forecasts of the volatility should all converge to the same normal level. Hence, the average of the instantaneous volatility~(\ref{instvoldef1}) should converge to the normal level as time tends to infinity, this is done, for instance, by Engle and Patton~\cite{engle-pat} who define the normal level of volatility as the stationary average of the square of instantaneous volatility~(\ref{instvoldef1}):
$$
\lim_{t\rightarrow \infty} \average{\frac{dX(t)^2}{dt}\Biggl|\sigma_0}\equiv \mbox{ normal level of volatility}.
$$
Note that the limit over time $t$ indicates that process has begun in the infinite past and, therefore, the volatility process is in the stationary state.

Our O-U volatility process appears to be an adequate candidate for describing this effect. Let us show this. From Eq.~(\ref{dx}) and taking in to account the independence of $\sigma(t)$ and $dW_1(t)$ and that $\average{dW_1^2}=dt$, we get
$$
\lim_{t\rightarrow \infty} \average{dX(t)^2|\sigma_0}= \lim_{t\rightarrow \infty} \average{\sigma^2(t)\Bigl|\sigma_0}dt,
$$
but the limit $t\rightarrow \infty$ indicates that volatility has reached the stationary state. Hence, the second moment of sigma is given by Eq.~(\ref{statave1}), and therefore
\begin{equation}
\mbox{O-U normal level of volatility}=\theta^2+\frac{k^2}{2\alpha}.
\label{normvol}
\end{equation}
Observe that the O-U volatility has a constant and non zero normal level of volatility and this is in accordance with the observed mean-reverting property mentioned above.

Moreover, the average over $\sigma^2$ given by Eq.~(\ref{sigma}), when the volatility is not yet in the stationary state, is
$$
\average{\sigma^2(t)|\sigma_0,0}=\left[\sigma_0 e^{-\alpha t}+
\theta(1-e^{-\alpha t})\right]^2+\frac{k^2}{2\alpha}\left(1-e^{-2\alpha t}\right).
$$ 
We observe that this average quickly tends to the normal level~(\ref{normvol}) as $\alpha t$ increases.  This is the reason why the magnitude of $\alpha$ allows us to classify the SV models into: (i) fast mean reverting processes when $1/\alpha\ll t$, and (ii) slow mean reverting processes when $1/\alpha \gg t$.

\section{The leverage effect\label{sec:lev}}
\noindent
It is commonly known that positive or negative sudden changes in the return have not the same impact on the volatility. Fisher Black in 1976~\cite{black76} was the first to find empirical evidence on this and observed that the volatility is negatively correlated with return variations. A qualitative explanation of this effect is that a fall in the stock prices implies an increase of the leverage of companies, which in turn entails more uncertainty and hence higher volatility. Nevertheless, it has also been argued that the leverage alone is too small to explain empirical asymmetries in prices~\cite{ghys96}. Another possible explanation is that news on any increase of volatility reduce the demand of stock shares because of investor's risk aversion. The consequent decline in stock prices is followed by an increment of the volatility as initially forecasted by news, and so on~\cite{ghys96}.

Although the mechanism still lacks of a clear explanation, the leverage effect denomination indicates this negative correlation. To our knowledge, this effect has only been studied in a qualitative manner until very recently when Bouchaud {\it etal.}~\cite{bouchPRL} have performed a complete empirical analysis containing new important information on this issue. 

Following Bouchaud {\it etal.}~\cite{bouchPRL}, we quantify the leverage effect by means of the leverage correlation function that we define in the form
\begin{equation}
{\cal L}(\tau)\equiv
\frac{\average{dX(t+\tau)^2dX(t)}}{\var[dX(t)]^2},
\label{leverage}
\end{equation}
where $X(t)$ is the zero-mean return defined in Eq.~(\ref{dXdef}). Bouchaud {\it etal.}~\cite{bouchPRL} have analyzed a large amount of daily relative changes for either market indices and stock share prices and find that
\begin{equation}
{\cal L}(\tau)=
\left\{
\begin{array}{ll}
-A e^{-b\tau} & \mbox{if } \tau>0, \\
0 & \mbox{if } \tau<0,
\end{array}
\right.
\label{bouchaudlev}
\end{equation}
$(A>0)$. Hence, there is a negative correlation with an exponential time decay between future volatility and past returns changes but no correlation is found between past volatility and future price changes. In this way, they provide a sort of causality to the leverage effect which, to our knowledge, has never been previously mentioned in the literature~\cite{fouquellibre,bek2000}.

\begin{figure}
\vspace*{13pt}
\centerline{\psfig{file=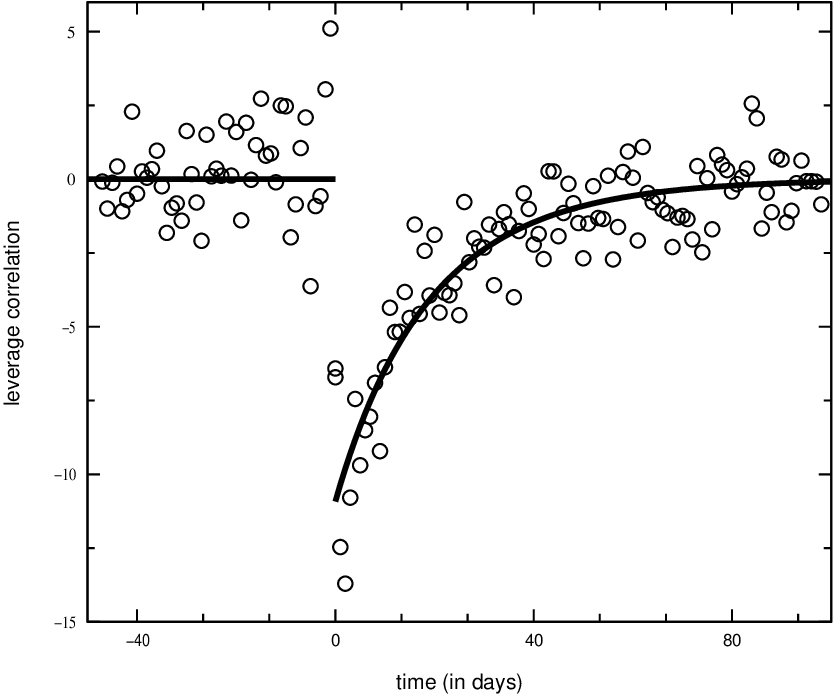,width=12cm,angle=-90}}
\vspace*{13pt}
\fcaption{The leverage effect in the Dow-Jones index. We plot the leverage function ${\cal L}(\tau)$ for the Dow-Jones index from 1900 until 2000. We see that there exists a non-negligible correlation when $\tau>0$ and negligible when $\tau < 0$. Observe that correlation strongly fluctuates when $-3<\tau<2$. We also plot a fit in solid lines with our O-U SV leverage function~(\ref{levou}). This fit helps us to estimate $\alpha$ and $\rho$ (see Section~\ref{sec:est} and Table~\ref{levestim} for more details).}
\label{djleverage}
\end{figure}

Let us see how our correlated O-U SV model is able to exactly reproduce this result. In effect, in the Appendix A we show that 
\begin{equation}
\var[dX(t)]=\theta^2(1+\nu^2)dt 
\label{var}
\end{equation}
where
\begin{equation}
\nu^2\equiv k^2/(2\alpha\theta^2),
\label{svnu}
\end{equation}
is the intensity of volatility fluctuations compared to the expected volatility
(see Eq.~(\ref{statave1})). On the other hand we prove in Appendix A that
\begin{equation}
\average{dX(t+\tau)^2dX(t)} = 
\left\{
\begin{array}{ll} 2\rho k \  \theta^2\left[1+\nu^2 e^{-\alpha\tau}\right]\displaystyle{e^{-\alpha \tau}} dt^2 & \mbox{if } \tau>0,
\\[0.2cm]
0 & \mbox{if } \tau<0.
\end{array}
\right.
\label{l}
\end{equation}
Hence the leverage correlation function~(\ref{leverage}) is
\begin{equation}
{\cal L}(\tau)=2\rho\left[\frac{\nu\sqrt{2\alpha}\left(1+\nu^2e^{-\alpha\tau}\right)}
{(1+\nu^2)^2\theta}\right] e^{-\alpha \tau} \qquad \mbox{for } \tau>0,
\label{levou}
\end{equation}
and
\begin{equation}
{\cal L}(\tau)=0 \qquad \mbox{for }\tau<0.
\label{L-}
\end{equation} 
We observe that sign of the leverage function is solely determined by the sign of $\rho$. Therefore, the O-U SV model is able to reproduce the empirically observed leverage correlation. In Fig.~\ref{djleverage} we show the leverage effect for the Dow-Jones index (1900-2000) and plot the leverage function given by Eq.~(\ref{levou}).

\section{Forecast evaluation\label{sec:est}}
\noindent
Any acceptable market model is required to ``forecast" the dynamics of the market. In other words, the model must be able to reproduce the market behavior and have an easy and systematic methodology for estimating its parameters. In our model these parameters are $\rho$, $k$, $\theta$, and $\alpha$~. Fouque {\it etal.}~\cite{fouque2000} for the same model only estimate two of these parameters ($k$ and $\theta$) from the empirical second and fourth moment of daily normalized stock changes. Unfortunately they cannot give a clear estimation of the volatility auto-correlation time $1/\alpha$ and of the magnitude of $\rho$ although it was already known that $\rho$ has to be negative in order to reproduce the desired skewness. Let us now see that with the help of the leverage effect we can completely estimate all parameters of the model. 

We will perform the empirical estimation of the Dow-Jones index daily return changes by approximating $dX$ by $\Delta X$, {\it i.e.}, 
$$ 
dX(t)\simeq X(t+\Delta t)-X(t),
$$ 
where $\Delta t=1$ day. From Eqs.~(\ref{var}) and~(\ref{variance2}) of Appendix A, we have
$$
\var[\Delta X]=\theta^2(1+\nu^2) \Delta t,
\qquad
\var[\Delta X^2]=2\theta^4\left[4(1+\nu^2)^2-3\right] \Delta t^2,
$$
where $\nu^2=k^2/(2\alpha\theta^2)$. From this we get
\begin{equation}
\frac{1}{(1+\nu^2)^2}=\frac43-\frac16 \frac{\var[\Delta X^2]}{\var[\Delta X]^2}.
\label{varestim}
\end{equation}
Hence, we can estimate the value of $\nu^2$ once we know the empirical values of these variances. The Dow-Jones index proportionates the daily variances of $\Delta X$ and $\Delta X^2$ and, subsequently, gives an estimated value for $\nu^2$. Afterwards, $\theta$ is estimated with the knowledge of $\nu^2$ and the empirical $\var[\Delta X]$. In Table~\ref{momestim}, we briefly report these operations and give the corresponding estimation of $\nu^2$ and $\theta^2$ for the Dow-Jones index time-series from 1900 until 2000.

Since $(1+\nu^2)^2$ is always positive and $(1+\nu^2)^2\geq 1$, we see from Eq.~(\ref{varestim}) that 
$$
2\leq\frac{\var[\Delta X^2]}{\var[\Delta X]^2}<8,
$$
and the kurtosis, $\gamma_2=\var[\Delta X^2]/\var[\Delta X]^2-2$, 
has the bounds
\begin{equation}
0\leq\gamma_2<6,
\label{bounds}
\end{equation}
showing that the model is never platykurtic. For the Dow-Jones index $\gamma_2=1.72$ and which is consistent with requirement~(\ref{bounds}). However, as Cont reports~\cite{cont2001}, there exists other markets or even intraday tick data possessing a higher kurtosis outside inequality~(\ref{bounds}).

\begin{table}[tbp]
\tcaption{The O-U SV estimation from the return variances. We estimate the parameters of the model from Dow-Jones historical daily returns from 1900 to 2000. We take the variances given by Eq.~(\ref{var}) and use the identity~(\ref{varestim}) for deriving the estimated quantities $\nu^2$ and $\theta$.}

\vspace*{13pt}
\begin{center}
\begin{tabular}{lcc}
\hline
\hline
\\
{\footnotesize Estimators} & {\footnotesize Dow-Jones daily return} & {\footnotesize Theoretical values} \\
\cline{1-3}
\\
$\var[\Delta X(t)]$ & $1.68 \times 10^{-4}$ & $\theta^2 \ (1+\nu^2) \Delta t$
\\ \\
$\var\left[\Delta X(t)^2\right]$ & $10.5 \times 10^{-8}$ & $2\theta^4\left[4(\nu^2+1)^2-3\right] \Delta t^2$ \\ \\ 
{\footnotesize Parameter estimation} & $\nu^2=0.18$ \\ \\ 
& $\theta=18.9 \% \ \mbox{year}^{-1/2}$ \\ \\
\hline \hline
\end{tabular}
\end{center}
\label{momestim}
\vspace*{13pt}
\end{table}

\begin{table}[tbp]
\tcaption{The O-U SV estimation from the leverage. We estimate the parameters $\rho$, $\alpha$ and $k$ from the fit of the leverage correlation derived from the Dow-Jones stock index data plotted in Fig.~\ref{djleverage}. For doing this, we take the $\nu^2$ estimation given by Table~\ref{momestim} and assume that the leverage function is given by Eq.~(\ref{estimrho}). Observe that magnitudes ${\cal L}(0^+)$ and $\alpha$ estimated from the Dow-Jones index are of the same order as those given by Bouchaud {\it etal.}~\cite{bouchPRL} for a combination of several stock indices.}

\vspace*{13pt}
\begin{center}
\begin{tabular}{lc}
\hline \hline \\
{\footnotesize Estimators} & {\footnotesize Dow-Jones data estimation} \\
\cline{1-2}
\\
${\cal L}(0^+)$ & -12.5 \\ \\
$\alpha$ & 0.05 $\mbox{day}^{-1}$ \\ \\
$1/\alpha$ & 19.6 days \\ \\
$\rho$ & -0.58 \\ \\
$k=\sqrt{2\alpha \nu^2 \theta^2}$ & $1.4 \times 10^{-3}\mbox{days}^{-1}$
\\
\\
\hline \hline
\end{tabular}
\end{center}
\label{levestim}
\vspace*{13pt}
\end{table}

The leverage effect provides the way for estimating the correlation coefficient $\rho$ and the characteristic time $1/\alpha$. Indeed, the best fit of the leverage function~(\ref{levou}) to the Dow-Jones daily data (Fig.~\ref{djleverage}) gives a characteristic time decay $1/\alpha\simeq 20$ days. We thus compare the empirical leverage with our theoretical leverage~(\ref{levou}) when $\tau\rightarrow 0^+$,
\begin{equation}
{\cal L}(0^+)=2\rho\frac{\nu\sqrt{2\alpha}}{(1+\nu^2)\theta},
\label{estimrho}
\end{equation}
and get an estimation of $\rho$ once we know $\alpha=0.04 \mbox{ days}^{-1}$ and $\nu^2=0.18$. Finally, we can derive $k$ from the definition given by Eq.~(\ref{svnu}) (see Table~\ref{momestim}). All these operations are summarized in Table~\ref{levestim}.

\begin{figure}[htbp]
\begin{center}
\end{center}
\centerline{\psfig{file=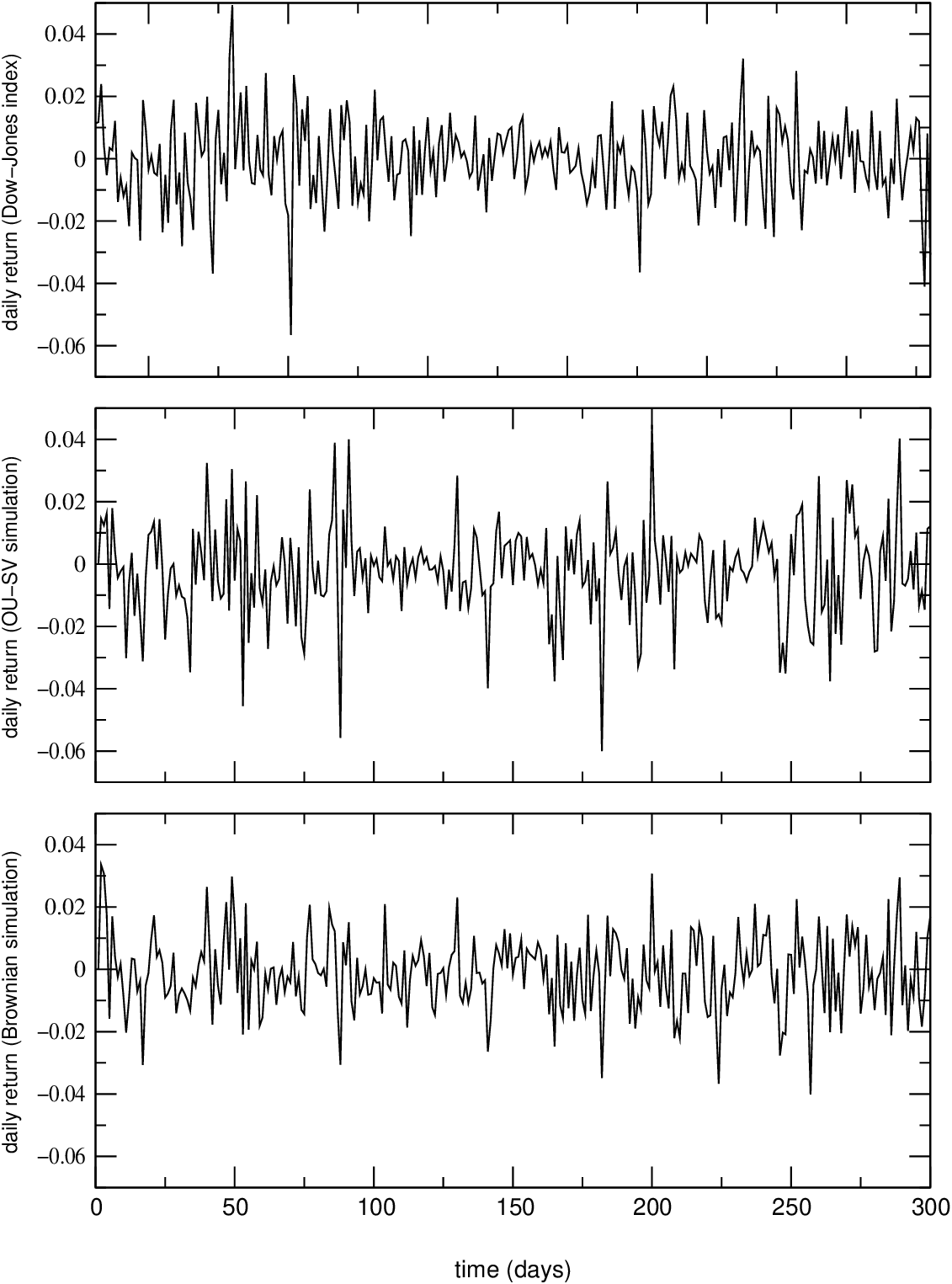,width=12cm}}
\vspace*{13pt}
\fcaption{Path simulation and Dow-Jones historical time-series. We show a Dow-Jones daily returns sample path (top figure), the O-U SV process simulation, and the geometric Brownian  process (with constant volatility) simulation (bottom figure). For the OU-SV process, we have taken the parameters given by Tables~\ref{momestim} and~\ref{levestim}. For the ordinary geometric Brownian motion, $\sigma^2=\var[\Delta X]/\Delta t$ and this variance is given in Table~\ref{momestim}. The dynamics is traced over approximately a trading year (the empirical path approximately corresponds to 1999 trading year).}
\label{comp}
\end{figure}

In Fig.~\ref{comp} we simulate the O-U SV resulting process with the parameters estimated above. We follow the random dynamics for $\Delta X(t)$ and compare it with the empirical Dow-Jones time series during approximately one trading year. We also simulate the ordinary geometric Brownian motion, {\it i.e.}, assuming a constant volatility whose value is directly estimated from data as in Table~\ref{momestim}. We clearly see in Fig.~\ref{comp} that the ordinary geometric Brownian model cannot describe neither largest nor smallest fluctuations of daily returns. Hence, in comparison with the ordinary Brownian modelisation, we conclude that the SV model describes a very similar trajectory than that of the Dow-Jones.  This is quite remarkable, because we have simulated last year's trajectory using all past data of the Dow-Jones index with almost equal results to the actual case. This, in turn, shows the stability of parameters. We may thus say that the model is fairly useful for ``predicting" the stock dynamics (at least for one year period) using market history.

\section{The probability distribution\label{sec:charou}}
\noindent
We will now obtain the probability distribution of the model. This problem has been recently addressed by Sh\"obel and Zhu~\cite{shobel} who, using the Feynmann-Kac functional, end up with an expression for the two-dimensional characteristic function of the joint process $(R(t),\sigma(t))$. Here we take a different path that, besides being simpler, allows us to get an analytical expression of the return characteristic function, which has more practical interest than the joint density. The analysis can be done in terms of the return $R(t)=\ln S(t)/S_0$ but we prefer to deal with the zero-mean return $X(t)$ since, although the calculation is basically identical, the expressions derived are shorter and handier.

\subsection{The characteristic function}
\noindent
Let $p_2(x,\sigma,t|x_0,\sigma_0,t_0)$ be the joint probability density of the two-dimensional diffusion process $(X(t),\sigma(t))$ described by the pair of SDE's given by Eqs.~(\ref{dx}) and~(\ref{dsigma}). This density obeys the following backward Fokker-Planck equation~\cite{gardiner}:
\begin{equation}
\frac{\partial p_2}{\partial t_0}=\alpha(\sigma_0-\theta) \frac{\partial p_2}{\partial \sigma_0}-\frac12 \sigma_0^2 \frac{\partial^2 p_2}{\partial x_0^2}-\rho k \sigma_0 \frac{\partial^2 p_2}{\partial \sigma_0 \partial x_0}-\frac12 k^2 \frac{\partial^2 p_2}{\partial \sigma_0^2},
\label{partdif}
\end{equation}
with final condition
\begin{equation}
p_{2}(x,\sigma,t|x_0,\sigma_0,t)=\delta(x-x_0) \ \delta(\sigma-\sigma_0).
\label{finalcond}
\end{equation}
Before proceeding further we note that, since all coefficients in Eq.~(\ref{partdif}) are independent of $x,x_0,t$ and $t_0$ and the final condition only depends on the differences $x-x_0$ and $t-t_0$, then $(X(t),\sigma(t))$ is an homogenous process in time and return. Therefore, 
\begin{equation}
p_{2}(x,\sigma,t|x_0,\sigma_0,t_0)=p_{2}(x-x_0,\sigma,t-t_0|\sigma_0).
\label{hom}
\end{equation}
Moreover, one easily sees that the marginal density of the return,
$$
p_X(x-x_0,t-t_0|\sigma_0)=\int_{-\infty}^{\infty} p_{2}(x-x_0,\sigma,t-t_0|\sigma_0) d\sigma,
$$
also obeys the same partial differential equation than $p_2$, Eq.~(\ref{partdif}), which due to homogeneity can be written in the form\footnote{Note that Eq.~(\ref{partdif}) is the backward Fokker-Planck equation while Eq.~(\ref{back1}) is a forward equation for $x$ and $t$ but not for $\sigma$.}
\begin{equation}
\frac{\partial p_{X}}{\partial t}=-\alpha(\sigma_0-\theta)\frac{\partial p_{X}}{\partial\sigma_0}+\frac12 \sigma_0^2 \frac{\partial^2 p_{X}}{\partial x^2}-\rho k \sigma_0 \frac{\partial^2 p_{X}}{\partial \sigma_0 \partial x}+\frac12 k^2 \frac{\partial^2 p_{X}}{\partial \sigma_0^2},
\label{back1}
\end{equation}
where $p_X=p_X(x-x_0,t-t_0|\sigma_0)$ and the initial condition is
$$
p_{X}(x-x_0,0|\sigma_0)=\delta(x-x_0).
$$
Partial differential equation~(\ref{back1}) is the starting point of our analysis. Again, due to homogeneity we can assume, without loss of generality that $x_0=0$ and $t_0=0$. Observe that 
Eq.~(\ref{back1}) is quite involved because of the correlation between the volatility and the return, {\it i.e.}, because of the crossed derivative term. Therefore, there seems to be a tremendous task if one tries to attack the problem directly from Eq.~(\ref{back1}). However, Fourier analysis proportionates the necessary tools to obtain an analytic solution of the problem. Solution that is expressed in terms of the return characteristic function (cf) defined by
$$
\varphi_X(\omega,t|\sigma_0)=\int_{-\infty}^{\infty} e^{i\omega x} p_X(x,t|\sigma_0) dx.
$$
This is done in Appendix B where we prove that
\begin{equation}
\varphi_X(\omega,t|\sigma_0)=\exp[-A(\omega,t)\sigma_0^2-B(\omega,t)\sigma_0-C(\omega,t)],
\label{char1}
\end{equation}
where
\begin{eqnarray}
&&A(\omega,t)=\frac{\omega^2}{2} \left(\frac{\sinh\eta t}{\eta \cosh\eta t+\zeta \sinh\eta t}\right),
\label{svA}
\\
&&B(\omega,t)=\frac{\omega^2 \alpha \theta}{\eta} \left(\frac{\cosh\eta t-1}{\eta \cosh\eta t+\zeta \sinh\eta t} \right),
\label{B}
\\
&&C(\omega,t)=\left[\frac{(\omega\alpha\theta)^2}{\eta^2}+i\omega\rho k-\alpha\right]t/2+\frac12 \ln\left(\cosh\eta t+\frac{\zeta}{\eta} \sinh\eta t\right)\nonumber \\
&&\qquad \qquad \qquad \qquad \qquad-\frac{(\omega\alpha\theta)^2}{2\eta^3} \left[\frac{2\zeta(\cosh\eta t-1)+\eta
\sinh\eta t}{\eta \cosh\eta t+\zeta \sinh\eta t}\right],
\label{C}
\end{eqnarray}
and
\begin{eqnarray}
\eta = \sqrt{\alpha^2-2i\rho k\alpha \omega+(1-\rho^2)k^2\omega^2}, \qquad 
\zeta =\alpha-i\omega\rho k.
\label{def}
\end{eqnarray}
Furthermore, we obtain the unconditional characteristic function, $\varphi_X(\omega,t)$, if we average over $\sigma_0$ which we assume is in the stationary regime. We thus write
\begin{equation}
\varphi_X(\omega,t)=\int \varphi_X(\omega,t|\sigma_0) p_{\rm st}(\sigma_0) d\sigma_0
\label{35}
\end{equation}
with the stationary pdf given by Eq.~(\ref{sigmapdfstat}):
\begin{equation}
p_{\rm st}(\sigma)=\frac{1}{\sqrt{\pi k^2/\alpha}}\ \exp\left[-\frac{\alpha(\sigma-\theta)^2}{k^2}\right].
\label{svp0}
\end{equation}
Then from Eqs.~(\ref{char1}) and (\ref{35})-(\ref{svp0}), we get
\begin{equation}
\varphi_X(\omega,t)=\frac{1}{\sqrt{1+k^2 A/\alpha}} \exp\left[-C+\frac{B^2k^2/\alpha-4\theta B-4\theta^2 A}{4(1+k^2 A/\alpha)}\right].
\label{char2}
\end{equation}
Note that this solution has the right limit when volatility is constant and non-random. Indeed, in such a case $k=0$ and from Eqs.~(\ref{svA})-(\ref{C}) and~(\ref{char2}) we have
\begin{equation}
\varphi_X(\omega,t)= e^{-\omega^2\theta^2 t/2},
\label{clt}
\end{equation}
which is the cf of the zero-mean return when $X(t)$ follows a one-dimensional diffusive process with constant volatility $\sigma=\theta$. Hence, solution~(\ref{char2}) appears to be consistent with the geometric Brownian motion model.

\subsection{Convergence to the Gaussian distribution}
\noindent
Let us see that, as $t\rightarrow\infty$, the marginal distribution~({char2}) approaches to the Gaussian density. In other words, we will prove a Central Limit Theorem for the model. The starting point of this analysis is Eq.~(\ref{char2}) that as $\alpha t\gg 1$ can be written in the following simpler form
\begin{equation}
\varphi_X(\omega,t)\sim\exp\left[-\left(\frac{\omega^2\alpha^2\theta^2}{\eta^2}+i\omega\rho k-\alpha+\eta\right)t/2\right],\qquad(\alpha t\gg 1).
\label{asin1}
\end{equation}
This is not a Gaussian distribution yet, since $\eta$ defined by Eq.~(\ref{def}) is an irrational function of $\omega$. We have to assume an extra requirement. Specifically, we suppose that 
\begin{equation}
\frac{k}{\alpha}\ll 1,
\label{asin2}
\end{equation}
which means that volatility is weakly random. Indeed, $k$ is the strength of the volatility driving 
noise (the so-called volatility of volatility) while $\alpha$ tell us how large is its deterministic drift   (see Eq.~(\ref{dsigma})). Therefore the ratio $k/\alpha$ measures in some way the degree of volatility randomness. Taking into account Eq.~(\ref{asin2}) we write 
(\cf Eq.~(\ref{def}))
$$
\eta=\alpha\left[1-i\omega\rho\frac{k}{\alpha}+\frac 12\omega^2\frac{k^2}{\alpha^2}+
\mbox{O}\left(\frac{k^3}{\alpha^3}\right)\right].
$$
Substituting this into Eq.~(\ref{asin1}) yields
\begin{equation}
\varphi_X(\omega,t)\sim\exp\left\{-\omega^2\left[1+\nu^2+\mbox{O}(k/\alpha)\right]\theta^2t/2\right\},
\qquad(\alpha t\gg 1),
\label{asin4}
\end{equation}
where $\nu^2\equiv k^2/2\alpha\theta^2$. The Gaussian density~(\ref{asin4}) proves the Central Limit Theorem in our case. 

\subsection{Cumulants}
\noindent
Cumulants are defined as follows
$$
\kappa_n\equiv\left.(-i)^n\ \frac{\partial^n}{\partial \omega^n}\ln[\varphi_X(\omega,t)]\right|_{\omega=0},
$$
and are very useful for deriving statistical properties of the model. For instance, the second cumulant reads
\begin{equation}
\kappa_2=\theta^2\left(1+\nu^2 \right)t,
\label{cum2}
\end{equation}
which results to be the integrated variance of $dX(t)$ (\cf Eq.~(\ref{variance2})). The third and fourth cumulants are respectively
\begin{equation}
\kappa_3=3\rho \frac{\theta^2 k}{\alpha^2}\left\{2 \left[\alpha t-\left(1-e^{-\alpha t}\right)\right]+\frac{\nu^2}{2}\left[2\alpha t-\left(1-e^{-2\alpha t}\right)\right]\right\},
\label{cum3}
\end{equation}
and 
\begin{eqnarray}
\kappa_4=k^2\theta^2\frac{3}{2\alpha^3}\Biggl\{
4\Bigl[2\alpha t+\alpha t\rho^2\left(6+4e^{-\alpha t}\right)-\left(2+12\rho^2\right)\left(1-e^{-\alpha t}\right)\nonumber\\ +\rho^2\left(1-e^{-2\alpha t}\right)\Bigr]
+\nu^2\left[2\alpha t+8\alpha t\rho^2\left(1+e^{-2\alpha t}\right)-\left(1-e^{-2\alpha t}\right)\left(1+8\rho^2\right)\right]\Biggr\}.
\label{cum4}
\end{eqnarray}
In terms of cumulants kurtosis is given by
\begin{equation}
\gamma_2\equiv \frac{\kappa_4}{\kappa_2^2},
\label{kurt}
\end{equation}
and measures the tails of the distribution compared to the Gaussian distribution. The kurtosis can be thus obtained using the second and fourth cumulants given by Eqs.~(\ref{cum2}) and~(\ref{cum4}). The resulting expression is very similar to that of the fourth cumulant~(\ref{cum4}) but with an extra constant factor. Its asymptotic limits are rather simple to derive and read
\begin{equation}
\gamma_2 \sim \frac{6\nu^2(\nu^2+2)}{(\nu^2+1)^2}\qquad(\alpha t\ll 1),
\label{t<1}
\end{equation}
and 
\begin{equation}
\gamma_2\sim \frac{6\nu^2[\nu^2(1+4\rho^2)+4(1+\rho^2)]}{(\nu^2+1)^2} \ \frac{1}{\alpha t}
\qquad(\alpha t\gg 1).
\label{t>1}
\end{equation}
From Eq.~(\ref{t<1}), we observe that even dealing with an infinitesimal time there will exist a non negligible kurtosis. Conversely, from Eq.~(\ref{t>1}) we see that kurtosis goes to zero as time increases and that the convergence is slow going as $1/t$. In addition, we observe that for short times kurtosis does not contain the correlation coefficient $\rho$ but in the long run a non zero $\rho$ magnifies the kurtosis of the distribution (\cf Eqs.~(\ref{t<1})-(\ref{t>1})). 

Let us now turn our attention to skewness $\gamma_1$ defined in terms of third and second cumulant as 
\begin{equation}
\gamma_1\equiv \frac{\kappa_3}{\kappa_2^{3/2}}
\label{ske}
\end{equation}
and that quantifies the byass in the return distribution. A negative skewness indicates that returns are more likely to decrease than to increase, while a positive skewness indicates a higher probability of a return raising than a decline. Empirical observations have found that financial markets have an slightly negative skewness~\cite{cont2001}. 

Similarly to the kurtosis derivation, the skewness defined by Eq.~(\ref{ske}) is obtained with Eqs.~(\ref{cum2})-(\ref{cum3}) and the resulting expression is very similar to that of Eq.~(\ref{cum3}) but with an extra constant factor. We limit ourselves to the asymptotic cases
\begin{equation}
\gamma_1\sim 3\rho \ \frac{\nu}{\sqrt{\nu^2+1}} \ \sqrt{2\alpha t}\qquad (\mbox{for } \alpha t\ll 1),
\label{t<1s}
\end{equation}
and 
\begin{equation}
\gamma_1\sim 6\rho \ \frac{\nu(\nu^2+2)}{(\nu^2+1)^{3/2}} \ \frac{1}{\sqrt{2\alpha t}} 
\qquad(\mbox{for } \alpha t\gg 1).
\label{t>1s}
\end{equation}
From these equations we see that both at short and long times, skewness vanishes. However, it decreases very slowly, more slowly than kurtosis (compare Eqs.~(\ref{t>1}) and~(\ref{t>1s})). Finally, we note that skewness is proportional to $\rho$ and, in consequence, the sign of $\rho$ not only determines the leverage correlation sign but also the skewness sign. Empirical observations of leverage and skewness indicate that $\rho$ must be negative~\cite{cont2001}.

\subsection{Tails}
\noindent
It is well established that distribution of prices have heavy tails. There exists several empirical studies quantifying this fact (see for instance Mantegna and Stanley~\cite{mante95} or Plerou {\it etal.}~\cite{plerou99}. Let us now study the existence of fat tails in our SV model. Recall first that for long times, $\alpha t\gg 1$, the probability distribution is practically Gaussian and there is no fat tail to look for. Therefore, we will search for heavy tails at small to moderate times, {\it i.e.}, when Central Limit Theorem is not applicable.

The tails of the distribution are determined by the shape of the density function $p_X(x,t)$ when $x$ is large. A well known fact from Fourier analysis is that the large $x$ behavior of $p_X(x,t)$ is given by the small $\omega$ behavior of its characteristic function $\varphi_X(\omega,t)$~\cite{weissllibre}. Therefore, tails are derived from the characteristic function~(\ref{char2}) by only keeping the first orders in $\omega$.

When $\omega$ is small but time is not too long ({\it i.e.}, $\alpha t\sim 1$) the expressions for $A(\omega,t)$, $B(\omega,t)$ and $C(\omega,t)$ are approximately given by (\cf Eqs.~(\ref{svA})-(\ref{C}))
$$
A(\omega,t)\sim \frac{\omega^2}{4\alpha}(1-e^{-2\alpha t}),\qquad 
B(\omega,t)\sim \frac{\theta\omega^2}{2\alpha}(1-e^{-\alpha t})^2,
$$
and
\begin{eqnarray*}
C(\omega,t)\sim(\omega^2\theta^2+i\omega\rho k-\alpha)t/2
-(\theta^2/4\alpha)[2(1-e^{-\alpha t})^2+1-e^{-2\alpha t}]\omega^2\\
+\frac12 \ln[1-i\rho kt\omega+(k^2/4\alpha^2)(2\alpha t-2\alpha^2t^2\rho^2-1+e^{-2\alpha t})\omega^2].
\end{eqnarray*}
Thus from Eq.~(\ref{char2}) we have
$$
\varphi_X(\omega,t)\sim\frac{[1+(k^2/4\alpha^2)(1-e^{-2\alpha t})]^{-1/2}
\exp(-\omega^2\theta^2t/2-i\omega\rho kt/2)}
{[1-i\rho kt\omega+(k^2/4\alpha^2)(2\alpha t-2\alpha^2t^2\rho^2-1+e^{-2\alpha t})\omega^2]^{1/2}}.
$$
Again, taking into account that $\omega$ is small and $t$ is moderate we get
\begin{equation}
\varphi_X(\omega,t)\sim\frac{1}{1-2ia(t)\omega+b(t)\omega^2}, \qquad(\omega\rightarrow 0),
\label{char2s}
\end{equation}
where
\begin{equation}
a(t)\equiv\rho kt/4, \qquad \mbox{and }\qquad 
b(t)\equiv k^2t(2-\rho^2\alpha t)/8\alpha.
\label{sva}
\end{equation}
The inverse Fourier transform for this asymptotic cf is
$$
p_X(x,t)\sim\frac{1}{\sqrt{a(t)^2+b(t)}} \exp\left[-\frac{1}{b(t)}
\left(\sqrt{a(t)^2+b(t)}|x|-a(t)x\right)\right],\quad(|x|\rightarrow\infty).
$$
Hence the tails of the zero-mean return have an asymmetric exponential decay given by
\begin{equation}
p_X(x,t)\sim
\frac{1}{\sqrt{a(t)^2+b(t)}}\exp\left[-\frac{1}{b(t)}\left(\sqrt{a(t)^2+b(t)}-a(t)\right)x\right]
\quad(x\rightarrow\infty),
\label{tail+}
\end{equation}
and
\begin{equation}p_X(x,t)\sim
\frac{1}{\sqrt{a(t)^2+b(t)}}\exp\left[\frac{1}{b(t)}\left(\sqrt{a(t)^2+b(t)}+a(t)\right)x\right],
\quad (x\rightarrow-\infty).
\label{tail-}
\end{equation}
Since $a(t)=\rho kt/4$, we see that the sign of $\rho$ will determine which is the heaviest tail. When $\rho$ is negative the fattest tail is the one representing losses and when $\rho>0$ the fattest tail corresponds to profits. If $\rho=0$ there is no difference between the two tails.

Finally, let us guess how the tails of price (not return) distribution are. We first recall that the asymptotic expressions (\ref{tail+})-(\ref{tail-}) refer to the marginal distribution of $X(t)$, that is, regardless the value of volatility at time $t$ and after averaging over the initial volatility. In order to obtain the asymptotic form of price distribution $p_S(S,t)$ out of the asymptotic form of $p_X(x,t)$ we must know what is the relation between $S(t)$ and $X(t)$. For the general case this relation is given by Eq.~(\ref{x(t)}) of Appendix A and, since it corresponds to the two-dimensional case when no average and marginal distribution have been performed, Eq.~(\ref{x(t)}) involves $X(t)$, $S(t)$ and $\sigma(t)$. We conjecture that if time is not too large $X(t)\sim\ln[S(t)/S_0]$. Therefore, for the price distribution $p_S(S,t)$ we have the following power laws:
\begin{equation}
p_S(S,t)\sim \frac{1}{S^{\nu_{-}(t)}}\quad (S\rightarrow 0),\qquad\mbox{and}\qquad
p_S(S,t)\sim \frac{1}{S^{\nu_{+}(t)}}\quad (S\rightarrow\infty)
\label{tailsS}
\end{equation}
where 
$$
\nu_{\pm}(t)=1+\frac{1}{b(t)}\left[\pm\sqrt{a(t)^2+b(t)}-a(t)\right].
$$

\section{Conclusions}
\noindent
The stochastic volatility (SV) models are a possible way out to the observed inconsistencies between the geometric Brownian model and real markets. The SV models, as their name indicates, assume the original log-Brownian model but with the volatility $\sigma$ being random. From 1987 on, there have appeared several works extending and refining stochastic volatility models but most of them are basically designed to reproduce empirical option prices. We have assumed that the volatility follows one of the simplest stochastic volatility models still showing mean reversion, {\it i.e.}, the Ornstein-Uhlenbeck process, but also allowing for correlations between the random fluctuations of the price and volatility processes.

Recent efforts in the study of the empirical statistical properties of the speculative markets has led to a list of stylized facts that all proposed model should accomplish. We have shown that the model is able to reproduce these facts. More specifically, the volatility process is mean reverting and have found that it is able to quantitatively reproduce the recently observed leverage effect. We have also estimated all parameters of the model and observed that a simulated path (using the estimated parameters) is very  similar to the sample path of the historical evolution of the Dow-Jones daily index (1900-2000). Finally, we have derived the characteristic function of the process, obtained its kurtosis and skewness, and observed the power-law decay for the tails of the price distribution. The results herein derived show that the SV models are good candidates for describing not only option prices but market dynamics as well.

We finally point out two further extensions of our work. First, the resulting characteristic function with all parameters already estimated can be used to derive option prices. Second, the observed volatility is not definitively attached to any specific model. There exists a large class of volatility processes and, since we now have a systematic way of estimating parameters of the process, a deeper analysis on which model gives a better explanation of the  stylized facts can also be very interesting. Other existent volatility models require a more involved mathematical analysis and it would be also interesting how far we can go using the methods presented in this paper. These two points are under present investigation.

\nonumsection{Acknowledgments}

\noindent
This work has been supported in part by Direcci\'on General de
Investigaci\'on under contract No. BFM2000-0795, by Generalitat de Catalunya under contract No. 2000 SGR-00023 and by Societat Catalana de F\'{\i}sica (Institut d'Estudis Catalans). 

\appendix{ A. The zero-mean return}

\noindent
The zero-mean return has been defined through its differential $dX$ by Eq.~(\ref{dXdef}). Let us first prove that $X(t)$ is explicitly given by
\begin{equation}
X(t)=\ln[S(t)/S_0]-\mu t - \frac{1}{2} \int_{0}^{t} \sigma^2(t') dt'.
\label{x(t)}
\end{equation}
Indeed, if we apply the It\^o lemma to Eq.~(\ref{x(t)}) and, as usual, keep orders smaller than $dt^{3/2}$ we have
$$
dX(t)=dS/S+\frac12 (dS/S)^2 - \mu dt -\frac12 \sigma^2 dt,
$$
but $dS/S=\mu dt+\sigma dW_1$ and $(dS/S)^2=\sigma^2 dt$, then we obtain Eq.~(\ref{dx}):
$$
dX=\sigma dW_1,
$$
and this proves the validity of Eq.~(\ref{x(t)}).

We will now derive several averages concerning $dX$. We know from Eq.~(\ref{mdx}) that
\begin{equation}
\average{dX(t)}=0.
\label{dx1}
\end{equation}
Again taking into account the independence of $\sigma(t)$ and $dW_1(t)$ we write
\begin{equation}
\average{dX^2}=\average{\sigma^2} \average{dW_1(t)^2},
\label{svdx2a}
\end{equation}
but $\average{dW_1^2}=dt$, and using Eq.~(\ref{statave1}) we have
\begin{equation}
\average{dX^2}=\left(\theta^2 +k^2/2\alpha\right)dt.
\label{svdx2}
\end{equation}
As to the fourth moment,
$$
\average{dX^4}=\average{\sigma^4}\average{dW_1^4},
$$
we note that $\average{dW_1^4}=3\average{dW_1^2}^2=3dt^2$ and we evaluate $\average{\sigma^4}$ using the stationary pdf~(\ref{svp0}). Hence
\begin{equation}
\average{dX^4}=3\left[3(k^2/2\alpha)^2+6 (k^2/2\alpha)\theta^2+\theta^4\right]dt^2.
\label{dx4} 
\end{equation}
The variances of $dX^2$ and $dX$ are obtained from Eqs.~(\ref{svdx2}) and~(\ref{dx4}) and read
\begin{eqnarray}
\var[dX(t)]&=&\left(\theta^2 +k^2/2\alpha\right)dt, \label{variance1} \\
\var[dX(t)^2]&=&2\left[\theta^4+4\left(k^2/2\alpha\right)^2+8\left(k^2/2\alpha\right)\theta^2\right]dt^2.
\label{variance2}
\end{eqnarray}

We finally derive the following correlation function:
$$
\average{dX(t+\tau)^2dX(t)} = \average{ \sigma(t) dW_1(t) \sigma(t+\tau)^2 dW_1(t+\tau)^2 }.
$$
Note that all variables are Gaussian which allows us to decompose the rhs of this equation into average pairs, taking also into account that $dW_1(t+\tau)^2=dt$ we can write
\begin{eqnarray*}
&&\average{dX(t+\tau)^2dX(t)} = \Bigl\{ 2 \average{ \sigma(t) \sigma(t+\tau) } \average{ \sigma(t+\tau) dW_1(t)} \nonumber \\
&& \qquad \qquad \qquad \qquad \qquad \qquad \qquad \qquad 
+\average{ \sigma(t+\tau)^2 } \average{ \sigma(t) dW_1(t)}\Bigr\}dt
\end{eqnarray*}
Combining this with Eqs.~(\ref{sigmadW1}) and~(\ref{mdx0}), we get
$$
\average{dX(t+\tau)^2dX(t)}=\left\{
\begin{array}{ll}2\rho k\average{ \sigma(t)\sigma(t+\tau)}\displaystyle{e^{-\alpha \tau}}dt^2 & 
\mbox{if } \tau>0,
\\
0 & \mbox{if } \tau<0;
\end{array}
\right.
$$
and the volatility correlation~(\ref{corrsig}) allows us to write
\begin{equation}
\average{dX(t+\tau)^2dX(t)} = 
\left\{
\begin{array}{ll} 2\rho k\ \left[\theta^2+(k^2/2\alpha) e^{-\alpha\tau}\right]
\displaystyle{e^{-\alpha \tau}} dt^2 & \mbox{if } \tau>0,
\\[0.2cm]
0 & \mbox{if } \tau<0.
\end{array}
\right.
\label{leverage2}
\end{equation}

\appendix{ B. The characteristic function of return}

\noindent
In this Appendix we will obtain the expression given by Eq.~(\ref{char1}) for the marginal characteristic function $\varphi_X(\omega,t|\sigma_0)$ of the two-dimensional diffusion process $(X,\sigma)$ whose joint density $p_2(x,\sigma,t|x_0,\sigma_0,t_0)$ is the solution to the final problem posed by Eqs.~(\ref{partdif})-(\ref{finalcond}). 

The marginal characteristic function (cf) of process $X(t)$,
$$
\varphi_X(\omega,t|\sigma_0)=\int_{-\infty}^{\infty}e^{i\omega x}p_X(x,t|\sigma_0)dx,
$$
allows us to write Eq.~(\ref{back1}) in the following simpler form:
\begin{equation}
\frac{\partial \varphi_{X}}{\partial t}=\frac12 k^2 \frac{\partial^2 \varphi_{X}}{\partial \sigma_0^2}+[i\omega \rho k\sigma_0-\alpha(\sigma_0-\theta)] \frac{\partial \varphi_{X}}{\partial \sigma_0}-\frac12 \sigma_0^2\omega^2 \varphi_{X}.
\label{back2}
\end{equation}
The initial condition is
\begin{equation}
\varphi_X(\omega,0|\sigma_0)=1.
\label{fc2}
\end{equation}

By direct inspection one can easily see that the solution to problem~(\ref{back2})-(\ref{fc2}) is 
\begin{equation}
\varphi_X(\omega,t|\sigma_0)=\exp[-A(\omega,t)\sigma_0^2-B(\omega,t)\sigma_0-C(\omega,t)]
\label{char1app}
\end{equation}
where functions $A(\omega,t), B(\omega,t)$ and $C(\omega,t)$ are the solution of the following set of  ordinary differential equations
\begin{eqnarray}
\dot{A}&=&-2k^2A^2-2(\alpha-i\omega\rho k)A+\omega^2/2\label{Aeq}\\
\dot{B}&=&-\left[2k^2A+(\alpha-i\omega\rho k)\right]B+2\alpha\theta A\label{Beq}\\
\dot{C}&=&k^2(A-B^2/2)+\alpha\theta B,\label{Ceq}
\end{eqnarray}
with initial conditions
\begin{equation}
A(\omega,0)=B(\omega,0)=C(\omega,0)=0.
\label{fchar}
\end{equation}

Note that Eq.~(\ref{Beq}) is a linear equation and that the rhs of Eq.~(\ref{Ceq}) does not contain $C(t)$. Therefore, their formal solutions are straightforward and read
\begin{equation}
B(t)=2\alpha\theta\int_0^tA(t')\exp\left[-(\alpha-i\omega\rho k)(t-t')-2k^2\int_{t'}^tA(t'')dt''\right]dt',
\label{BB}
\end{equation}
\begin{equation}
C(t)=k^2\int_0^t\left[A(t')-B^2(t')/2\right]dt'+\alpha\theta\int_{0}^tB(t')dt'.
\label{CC}
\end{equation}
On the other hand Eq.~(\ref{Aeq}) is more involved since it is a Ricatti equation. However, the definition of a new dependent variable
\begin{equation}
A=\frac{\dot{y}}{2k^2y}
\label{newvariable}
\end{equation}
turns Eq.~(\ref{Aeq}) into the following linear second-order equation with constant coefficients:
$$
\ddot{y}+2(\alpha-i\omega\rho k)\dot{y}-k^2\omega^2 y=0.
$$
The solution to this equation is 
$$
y(t)=C_1e^{\lambda_+ t}+C_2e^{\lambda_- t},
$$
where $C_{1,2}$ are arbitrary constants and 
$$
\lambda_{\pm}=\alpha-i\omega\rho k\pm\sqrt{(\alpha-i\omega\rho k)^2+k^2\omega^2}.
$$
Substituting this into Eq.~(\ref{newvariable}) yields
$$
A(t)=\frac{\lambda_+ e^{\lambda_+ t}+\lambda_-(C_2/C_1)e^{\lambda_- t}}
{2k^2[e^{\lambda_+ t}+(C_2/C_1)e^{\lambda_- t}]}.
$$
Now the initial condition $A(0)=0$ gives $C_2/C_1=-\lambda_+/\lambda_-$ and the substitution of $\lambda_{\pm}$ allows us to write $A(t)$ in the form given by Eq.~(\ref{svA}). Finally the substitution of Eq.~(\ref{svA}) into Eqs.~(\ref{BB})-(\ref{CC}) results in Eqs.~(\ref{B})-(\ref{C}).

\nonumsection{References}

\end{document}